\documentclass[aps,twocolumn]{revtex4}
\usepackage{setspace}
\usepackage{graphicx}         

\graphicspath{{images/}}

\begin{document}

\bibliographystyle{naturemag}

\title{Electromagnetically induced transparency with single atoms in a cavity}

\author{Martin M\"ucke\footnote{These authors contributed equally to this work}}
\author{Eden Figueroa$^*$\footnote{To whom correspondence should be addressed. Email: eden.figueroa@mpq.mpg.de}}
\author{Joerg Bochmann}
\author{Carolin Hahn}
\author{Karim Murr}
\author{Stephan Ritter}
\author{Celso J. Villas-Boas\footnote{Also with: Departamento de Fisica, Universidade Federal de S\~{a}o Carlos, 13565-905 S\~{a}o Carlos, SP, Brazil}} \author{Gerhard Rempe}

\affiliation{Max-Planck-Institut f\"ur Quantenoptik, Hans-Kopfermann-Str. 1, D-85748 Garching, Germany}

\maketitle

\textbf{Optical nonlinearities offer unique possibilities for the control of light with light. A prominent example is electromagnetically induced transparency (EIT) where the transmission of a probe beam through an optically dense medium is manipulated by means of a control beam \cite{Harris1997, Lukin2003, Fleischhauer2005}. Scaling such experiments into the quantum domain with one, or just a few particles of both light and matter will allow for the implementation of quantum computing protocols with atoms and photons \cite{Werner1999, Rebic1999, Bermel2006, Cardimona2009} or the realisation of strongly interacting photon gases exhibiting quantum phase transitions of light \cite{Hartmann2006, Greentree2006}. Reaching these aims is challenging and requires an enhanced matter-light interaction as provided by cavity quantum electrodynamics (QED) \cite{Imamoglu1997, Grangier1998, Imamoglu1998}. Here we demonstrate EIT with a single atom quasi-permanently trapped inside a high-finesse optical cavity. The atom acts as a quantum-optical transistor with the ability to coherently control \cite{Xu2008} the transmission of light through the cavity. We furthermore investigate the scaling of EIT when the atom number is increased one by one. The measured spectra are in excellent agreement with a theoretical model. Merging EIT with cavity QED and single quanta of matter is likely to become the cornerstone for novel applications, e.g. the dynamic control of the photon statistics of propagating light fields \cite{Rebic2002} or the engineering of Fock-state superpositions of flying light pulses \cite{Nikoghosyan2009}.}

Remarkable progress has been achieved towards the optical manipulation of light by means of single quantum emitters. First realisations of optical transistors operating on tightly focused laser beams have been demonstrated with individual molecules, quantum dots or nitrogen vacancies embedded in suitable host materials \cite{Santori2006, Xu2008, Xu2009, Hwang2009}. However, the weak light-matter coupling which has been reached in these experiments considerably limits the control capabilities. Moreover, increasing the number of particles is not a straightforward task, due to the difficulties in the preparation of identical quantum radiators.

\begin{figure}
  \includegraphics[keepaspectratio,width=1.0\columnwidth]{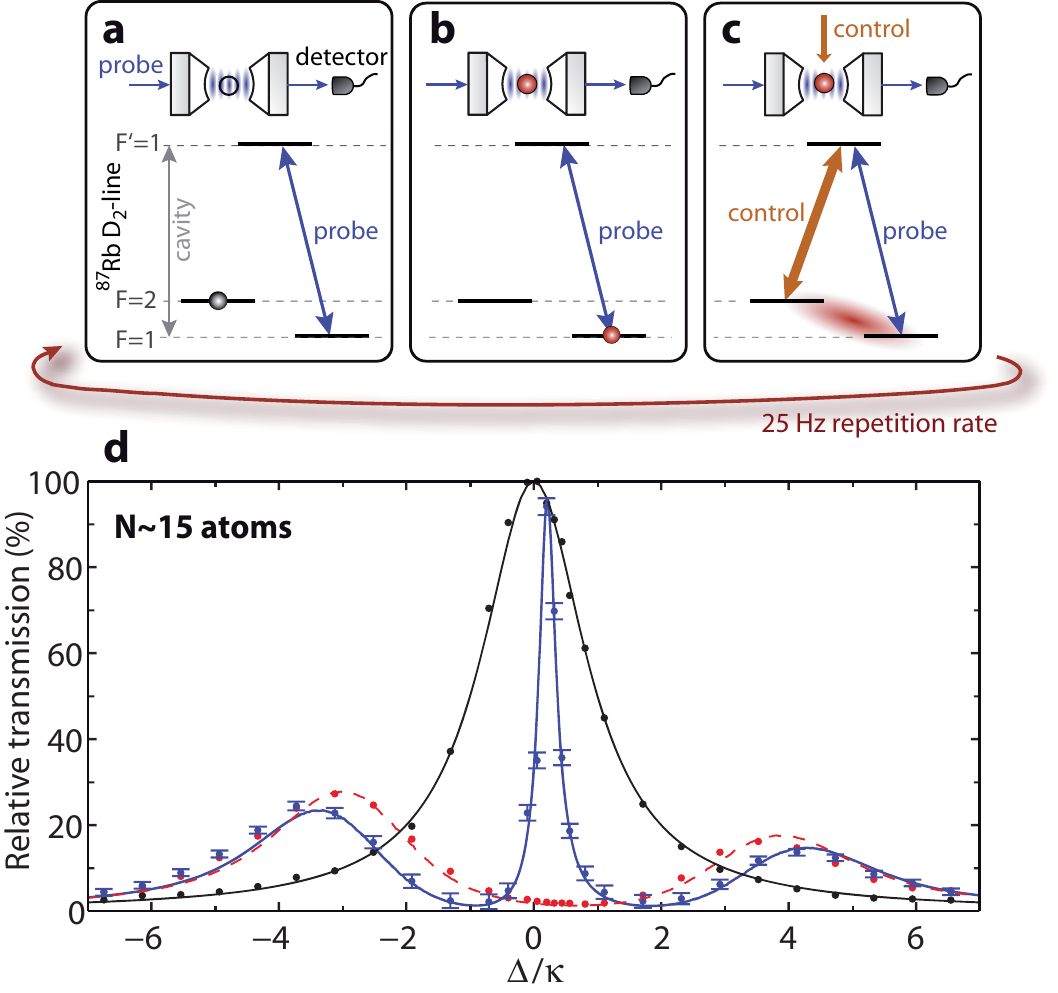}
  \caption{\textbf{Experimental protocol and cavity EIT.} Single $^{87}$Rb atoms are quasi-permanently trapped inside a high finesse optical cavity. The cavity is resonant with the atomic $F=1 \leftrightarrow F'=1$ transition at 780 nm wavelength. The transmission of the atom-cavity system is probed with a weak laser (probe laser-cavity detuning $\Delta$) for three physical conditions. \textbf{a}, With atoms shelved in the hyperfine state $F=2$, we record the empty cavity transmission as a reference (black data and curve in \textbf{d}). \textbf{b}, With atoms prepared in $F=1$, we realise a cavity QED situation and observe a spectrum exhibiting a vacuum-Rabi splitting (red data and curve in \textbf{d}). \textbf{c}, An additional laser is used to coherently control the optical properties of the atom-cavity system. \textbf{d}, Measured transmission spectra for on average 15 atoms coupled to the cavity. A narrow transmission window (full linewidth at half maximum $\approx$ 900 kHz) observed at the two-photon resonance in the cavity EIT scenario (blue data and theory curve) testifies to the existence of a coherent dark state. Experimental parameters: maximum intra-cavity photon number 0.02, control power 9 $\mu$W (equivalent Rabi frequency 1.3$\kappa$). Error bars are statistical and are omitted from the empty cavity and two-level measurement for clarity.
}\label{fig:Setup}
\end{figure}

With this backdrop, a promising avenue is to trap a register of atoms inside an optical cavity \cite{Nussmann2005, Khudaverdyan2009}. High-reflectivity mirrors increase the optical path length and can amplify the matter-light interaction into the strong coupling regime. Optical control has already been achieved in single-atom experiments, including the production of single photons with controlled waveform \cite{Kuhn2002, McKeever2004, Keller2004}, the generation of polarization-entangled photon pairs \cite{Wilk2007}, and the state transfer between a faint laser pulse and a single atom \cite{Boozer2007}. Incorporating EIT will boost the capabilities of cavity QED from the production of single photons towards the coherent manipulation of propagating quantum light fields. For a system with many individually addressable atoms, this will ultimately lead to the realisation of a quantum network \cite{Kimble2008}, where the generation, propagation and absorption of light is coherently controlled at the quantum level.

In this work, we coherently control the optical properties of a coupled atom-cavity system through the use of EIT. The heart of the apparatus consists of a high-finesse optical cavity with mirrors separated by 495$\mu$m, a TEM$_{00}$-mode waist of 30$\mu$m and a finesse of 56000. The cavity operates in the intermediate coupling regime with $(g_{0}, \kappa, \gamma) = 2\pi\times(4.5, 2.9, 3.0)$ MHz, where $g_{0}$ denotes the atom-cavity coupling constant at a field antinode for the $^{87}$Rb 5$S_{1/2}$ $F=1$ $\leftrightarrow$ 5$P_{3/2}$ $F'=1$ transition at 780\,nm, $\kappa$ is the cavity field decay rate and $\gamma$ the atomic polarisation decay rate. The atoms are trapped inside the cavity in a far-detuned standing-wave dipole trap. The cavity is stabilised to the $F=1 \leftrightarrow F'=1$ transition (see Methods). Light scattered during cooling intervals is used to obtain images of the trapped atoms with a CCD-camera. This allows the precise determination of the number and position of atoms inside the cavity mode during a given experimental run.

In order to demonstrate EIT in the regime of single atoms, we record transmission spectra of the atom-cavity system under three distinct physical conditions. The transmission is measured with a weak probe laser near resonant with the $F=1 \leftrightarrow F'=1$ transition applied along the cavity axis. In the first step of our experimental protocol, we shelve the atom in the hyperfine state $F=2$, therefore effectively decoupling it from the cavity, Fig. 1a. This yields an empty-cavity transmission spectrum used as a reference. In the second step, the atom is prepared in $F=1$, such that we realise the case of a two-level atom coupled to the cavity, Fig. 1b. In the third step, we apply an additional control laser transverse to the cavity axis and resonant with the $F=2 \leftrightarrow F'=1$ transition, hereafter named cavity EIT configuration, Fig. 1c. This forms a $\Lambda$-level scheme suitable for the generation of a coherent dark state. The experimental protocol is continuously repeated at a 25 Hz rate while the probe laser frequency is shifted for every repetition cycle. Thus, we simultaneously measure the three transmission spectra for a given number of trapped atoms.

We introduce the main features of cavity EIT by means of transmission spectra obtained with on average 15 atoms trapped inside the cavity, Fig. 1d. The data points and theory curve given in black correspond to the Lorentzian transmission of the empty cavity. In contrast, the transmission spectrum for the two-state atoms coupled to the cavity (red data and dashed curve) displays the characteristic vacuum-Rabi splitting accompanied by a significant drop in the transmission at the empty cavity resonance (probe-cavity detuning $\Delta=0$). This spectrum is dramatically altered under the conditions of EIT (blue data and theory curve). First, we notice a frequency shift of the vacuum-Rabi resonances due to the ''dressing'' of the atom-cavity energy levels by the control laser light. Most important, however, is the observation of a narrow transmission window testifying to the existence of a coherent dark state \cite{Lukin_CEIT_theory}. We verified that this transmission window appears at the two-photon resonance and can be shifted by changing the control-laser frequency. We emphasize that these results present an implementation of cavity EIT in a completely new regime, where the number of atoms is many orders of magnitude smaller than in previous realisations with atomic ensembles \cite{Zhu_CEIT_exp_1,Xiao_CEIT_exp_2}. It is the enhanced matter-light coupling per atom inside the optical resonator that allows us to observe EIT with only a few atoms.

The frequency-dependent transmission of our cavity EIT system can be described using a semi-classical theory in the limit of weak probe fields and assuming that almost all the population is in the atomic level $F=1$. This yields the steady-state transmission $T$, normalized to the transmission of the empty, resonant cavity, given by

\begin{equation}
T = \frac{\kappa^2}{\left\vert \left(\Delta+i\kappa\right) -g^{2}N\chi\right\vert ^{2}}
\label{eq:Photonaverage}
\end{equation}

Here, $N$ is the number of atoms and $\chi(\delta,\Omega_{c})$ the susceptibility of an EIT medium in free space \cite{Figueroa2006} which depends, in particular, on the two-photon detuning $\delta$ and the control-laser Rabi frequency $\Omega_c$. Furthermore, $g$ represents an effective atom-cavity coupling with $g^2=\sum^N_{i=1}g_{i}^2 /N$, where $g_i$ is the atom-cavity coupling of atom $i$. Equation \ref{eq:Photonaverage} relates the familiar expression for EIT, $\chi(\delta,\Omega_{c})$, to cavity QED with a countable number of atoms. In practice, the transmission spectra are subject to parameter variations and technical noise. Residual atomic motion in the optical dipole trap results mostly in variations of the atom-cavity coupling. Magnetic field fluctuations on short time scales contribute to the ground-state decoherence rate (measured to be about 65 kHz). These effects are incorporated into our theoretical model by averaging over the respective parameter range.

\begin{figure}
  \includegraphics[keepaspectratio,width=1.0\columnwidth]{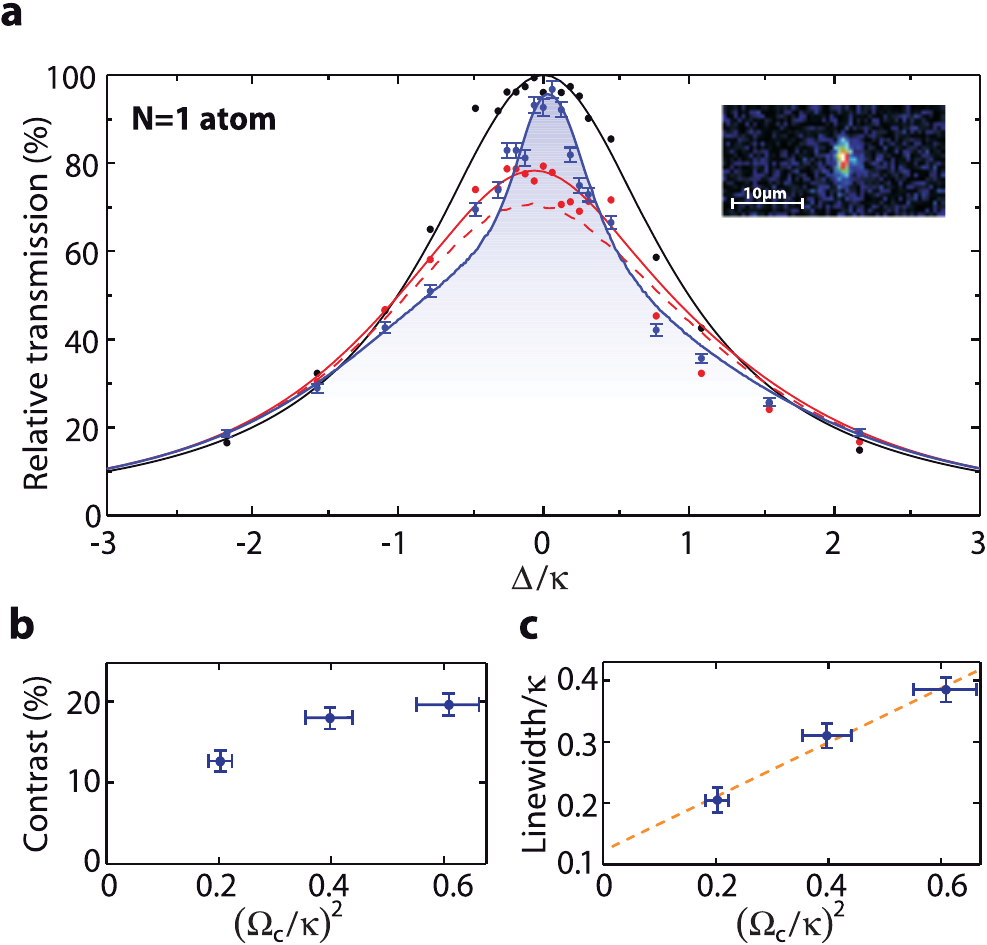}
  \caption{\textbf{Cavity EIT with a single atom.} Colour coding same as in Fig. 1d. \textbf{a}, Measured transmission spectra for exactly one atom coupled to the cavity and a control laser power of 3\,$\mu$W (equivalent Rabi frequency 0.78\,$\kappa$). EIT is observed with a maximum transparency of 96\,\% and a measured transmission contrast of 20\,\% with respect to the control laser switched off. The full linewidth at half maximum (see Methods) is $\approx$ 1.2 MHz. The inset shows a CCD camera image of a single atom trapped in the cavity (image size 33$\mu$m $\times$ 16$\mu$m). \textbf{b}, \textbf{c}, The linewidth and contrast of the single-atom transparency feature are tunable by means of the control laser power. The used values are (1, 2, 3)\,$\mu$W and correspond to Rabi frequencies (0.45, 0.63, 0.78)\,$\kappa$. The maximum intra-cavity photon number is 0.02. Error bars are statistical.}\label{fig:SA_CEIT}
\end{figure}

The interest in cavity EIT with a single atom has been put forward in several theory works over the past decade, especially in the context of nonlinear optics and quantum information \cite{Werner1999, Rebic1999, Bermel2006, Cardimona2009, Rebic2002, Nikoghosyan2009}. Its observation is presented in Fig. 2a. The data are an average over 169 complete spectra each obtained using exactly one trapped atom. For these measurements, the maximum intra-cavity photon number is $0.02$. For the case of a two-level atom (red data points), the on-resonance transmission is lowered but the vacuum-Rabi splitting is not resolved. This is a consequence of the atomic motion effectively reducing the coupling to an average value of approximately $0.4 g_0$. The weak probe field additionally induces a slow optical pumping to the $F=2$ hyperfine ground state during the probing period ($t=50\mu$s). This effect has been incorporated by solving the time-dependent master equation for the finite probing interval (red solid curve). The result is an increase in transmission, as compared to the steady state Eq.~\ref{eq:Photonaverage} (red dashed curve), which is in excellent agreement with the experimental data. By now turning on the control field, we observe cavity EIT with one atom (blue data points and curve). The transmitted spectrum is notably narrowed, and a nearly perfect transparency is obtained. In this respect, the system realises a quantum-optical transistor with an unprecedented on/off contrast of about $20\%$, admitting or rejecting the passage of probe photons through the cavity. Beyond that, the power of our approach lies in the preparation of the atom in a dark superposition of two states. This offers the coherent control of the optical properties of a single atom through external parameters. As an example, we investigate the dependence of the transparency window on the control laser power, Fig. 2b,c. The contrast shown in Fig. 2b increases with control-laser power, while its maximum value is bound by the difference between empty-cavity transmission and the two-level atom case. We find that the spectral width of the transparency window exhibits linear scaling with $\Omega_c^2$, Fig. 2c, which is in agreement with Eq.~\ref{eq:Photonaverage}. However, the determination of an absolute value of the EIT linewidth is obstructed by its non-Lorentzian shape due to the overlapping vacuum-Rabi resonances.

\begin{figure}
  \includegraphics[keepaspectratio,width=1.0\columnwidth]{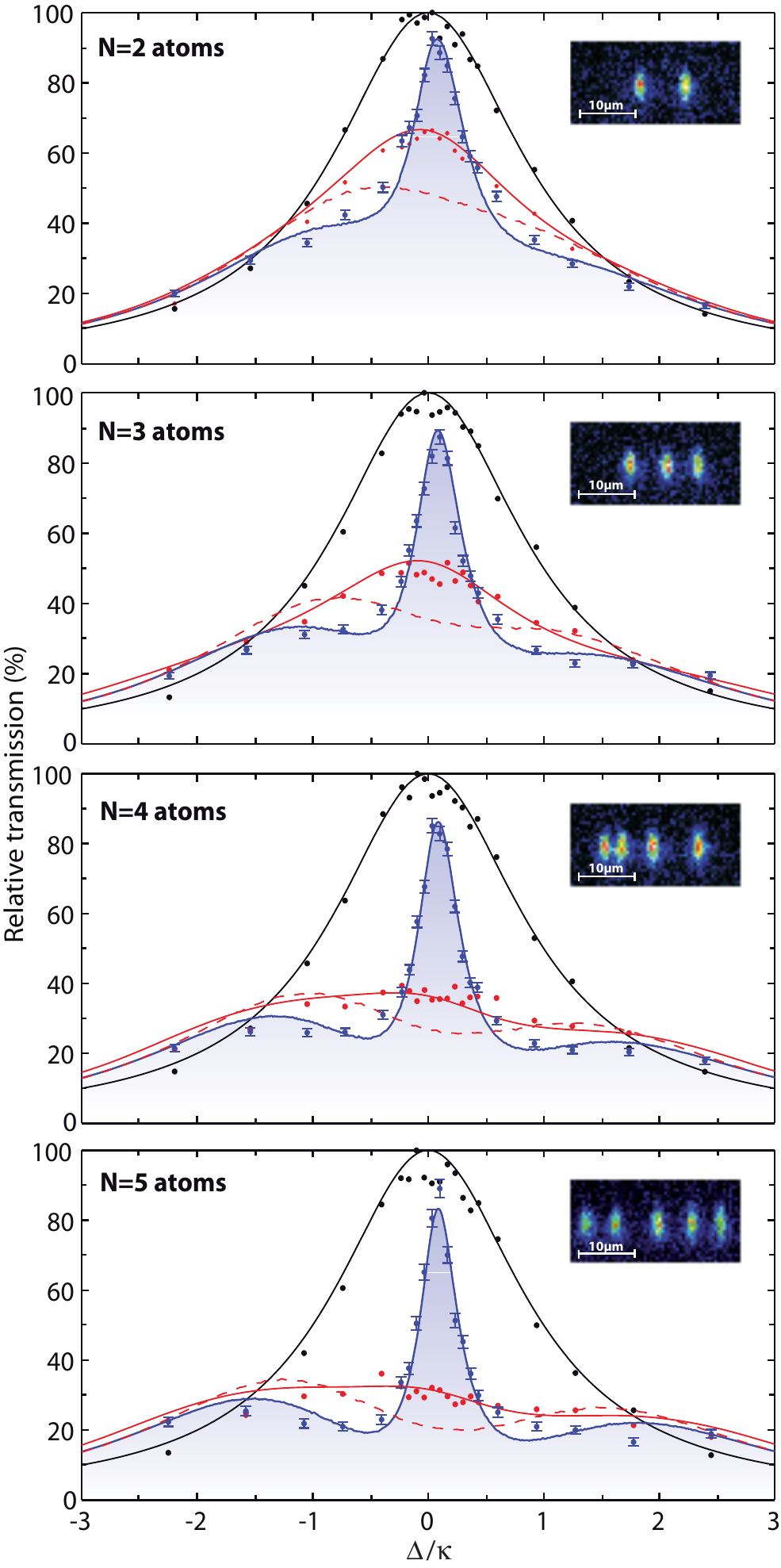}
  \caption{\textbf{Cavity EIT spectra for atoms by the number.} Changing the number of atoms enhances the visibility of the dark-state resonance due to improved contrast with respect to the measurements with two-level atoms coupled to the cavity. The vacuum-Rabi splitting starts being resolved at higher atom number. Colour coding and experimental parameters are the same as in Fig. 2a. The insets are CCD camera images of the trapped atoms, used to precisely determine their number and physical location in the cavity mode. Error bars are statistical and are omitted from the empty cavity and two-level measurement for clarity.} \label{fig:NA_CEIT}
\end{figure}

Our capability to determine the exact number of atoms in the cavity allows us to investigate the evolution of cavity EIT when adding atoms one by one. The experimental results are presented sequentially in Fig. 3 for atom numbers $N=2, 3, 4, 5$, with all other parameters identical to the $N=1$ atom case of Fig. 2a. The obtained spectra weakly reveal the vacuum-Rabi splitting, expected to scale with $\sqrt{N}$ for small probe-laser powers. The main effect of increasing the number of atoms is a lower transmission level at vanishing probe detuning ($\Delta=0$) (red curve). This uncovers the dark-state resonance, while a very high degree of transparency is maintained (solid blue). These observations are quantified by evaluating the scaling of transparency, contrast and linewidth as a function of atom number, see Fig. 4. The maximum achieved transparency decreases slightly from 96$\%$ ($N=1$) to 78$\%$ ($N=7$) as expected from Eq.~\ref{eq:Photonaverage}. Nevertheless, the measured on/off contrast at the two-photon resonance steadily increases from 20$\%$ ($N=1$) to 60$\%$ ($N=7$) largely due to the evolution of the two-level spectra. We note that the decrease in the maximum achieved transparency can be compensated by applying a stronger control field, as in the case of Fig. 1d. The dark-state linewidth (FWHM, see Methods) responds only weakly to the change in atom number, Fig. 4b. The linewidth decreases for $N\geq 3$. In the limit of $Ng^2/(\kappa\gamma) > 1$, here fulfilled for $N\geq 3$, and a negligible ground state decoherence rate, an approximation can be derived from Eq.~\ref{eq:Photonaverage} for which the linewidth scales as $\Omega_{c}^2/N$. This is consistent with our observations (dashed curve). However, for $N=1,2$ the determination of the EIT linewidth is difficult, as explained above. The rise of the two-level atom transmission (used as a reference to obtain the FWHM) compensates the expected broadening, resulting in a nearly constant linewidth for $N=1,2$.

\begin{figure}
  \includegraphics[keepaspectratio,width=1.0\columnwidth]{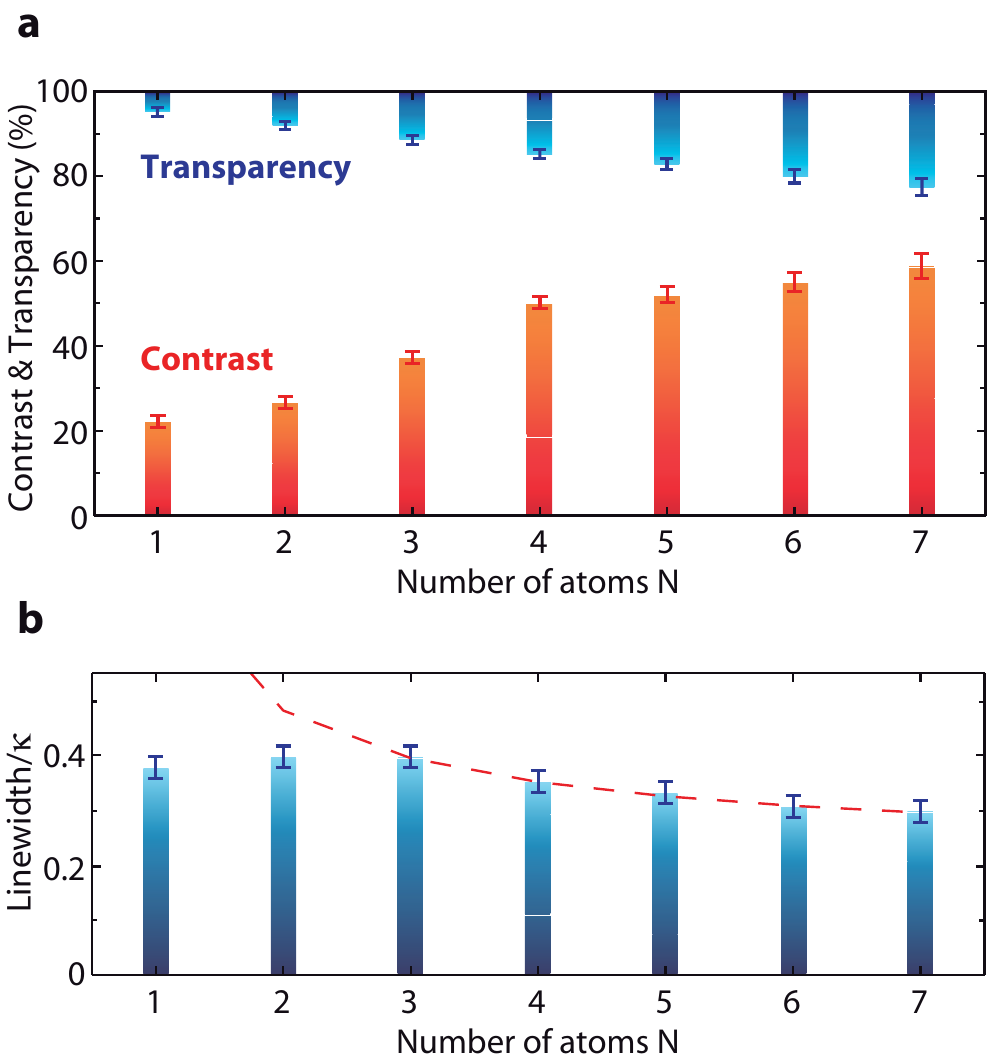}
  \caption{\textbf{Measured transparency, contrast and linewidth of cavity EIT with \emph{N}=1 to 7 atoms.} \textbf{a}, The maximum transparency (blue bars, top) decreases with the number of atoms from 96$\%$ ($N=1$) to 78$\%$ ($N=7$). Nevertheless, the on/off contrast (red bars, bottom) at the two-photon resonance steadily increases from 21$\%$ ($N=1$) to 60$\%$ ($N=7$) due to the reduction of transmission with control laser switched off. \textbf{b}, For $N\geq 3$, the cavity EIT linewidth decreases with the number of coupled atoms as $1/N$ (guide to the eye, dashed red curve). For $N=1,2$ the linewidth is nearly constant due to the interplay between the increase in the two-level atom transmission and the used definition of the linewidth (see Methods). Error bars are statistical.} \label{fig:contrast}
\end{figure}

Merging the ability to perform experiments with a controlled number of atoms \cite{Meschede2010} with EIT-based light storage \cite{Lukin2003} opens up new avenues towards the controlled manipulation of the Fock-state components of propagating light fields. As the number of photons which can be stored in a medium cannot exceed the number of atoms, it now seems possible to write a given number of photons into an atomic register and read out these photons later, while any excess photons would just pass the cavity. In this way a highly nonlinear beam splitter could be realised that subtracts a well-defined number of photons from a coherent input field with an initially Poissonian photon-number distribution.

In future experiments we plan to go beyond the coherent manipulation of the average light intensity and investigate the possibility to also control the photon statistics. The performance of such experiments would naturally benefit from a larger matter-light coupling constant, $g$, as can be achieved with a smaller cavity or with better atomic localisation. In fact, a moderate increase of $g$ by just a factor of three as achieved in other nonlinear single-atom cavity QED systems \cite{Schuster2008} would promote our present system well into the regime of strong photon-photon interactions. This could bring ideas like that of an EIT-controlled photon blockade in an atomic four-level system into reality, more than a decade after its proposal \cite{Imamoglu1997}. \\

\noindent \textbf{METHODS SUMMARY} \\
\footnotesize
\noindent \textbf{Atom trapping.} $^{87}$Rb atoms are loaded into the cavity mode from a magneto-optical trap (MOT) using a running-wave dipole trap ($\lambda$ = 1064 nm) with its focus halfway between the MOT and the cavity. When the atoms reach the cavity, the transfer beam is replaced by a standing-wave beam which provides strong confinement along its axis. The atoms are cooled by exposing them to retro-reflected laser beams incident perpendicularly to the cavity axis and near resonant with the $F=2 \leftrightarrow F'=3$ and $F=1 \leftrightarrow F'=2$ transitions. The cavity frequency is stabilised via a reference laser ($\lambda$ = 785 nm) close to the atomic $F=1 \leftrightarrow F'=1$ transition (bare atom-cavity detuning 2\,MHz, average AC-Stark shift 5\,MHz). \\
\noindent \textbf{Measurement sequence.} \emph{Empty cavity measurements}: Following 33 ms of optical cooling, the atoms are optically pumped to the $F=2$ ground state by applying a laser resonant to the $F=1 \leftrightarrow F'=2$ transition for 100 $\mu$s. After this, a weak probe laser (maximum intra-cavity photon number $\sim$ 0.02) near resonant to the $F=1 \leftrightarrow F'=1$ transition is applied along the cavity axis mode for 50$\mu$s. The cavity output is collected using a single-mode optical fibre and directed to a single-photon detection setup (total detection efficiency including propagation losses $\sim 30\,\%$). \emph{Two-level atom measurements}: After a cooling interval (3 ms), the atoms are now prepared in the $F=1$ ground state by simultaneously applying pumping beams resonant with the $F=2 \leftrightarrow F'=1$ and $F=2 \leftrightarrow F'=2$ transitions. The probe beam is applied again for 50 $\mu$s. The intensity of the standing-wave trapping field perpendicular to the cavity is reduced to minimize AC Stark shift variations of the atomic transitions ($\leq$ 5 MHz). \emph{Cavity EIT measurements}: After a cooling interval (3 ms), the atoms are prepared in the $F=1$ ground state. A control field resonant with the $F=2 \leftrightarrow F'=1$ transition perpendicular to the cavity axis is applied simultaneously with the probe. During the probe interval, the dipole-trap depth is reduced as for the two-level measurements. Probe and control laser fields maintain a fixed phase relation. \\
\noindent \textbf{Definition of cavity EIT linewidth.} We extract the Full Width at Half Maximum (FWHM) of the cavity EIT resonance from the experimental fit as $2\Delta_\mathrm{cEIT}$, in which $\Delta_\mathrm{cEIT}$ is the solution to the equation:
\[
T^\mathrm{fit}_\mathrm{cEIT}(\Delta=\Delta_\mathrm{cEIT}) = [T^\mathrm{fit}_\mathrm{cEIT}(\Delta=0) + T^\mathrm{fit}_\mathrm{2level}(\Delta=0)]/2
\]
\noindent where $T^\mathrm{fit}_\mathrm{cEIT}$ is represented by the solid blue line, and $T^\mathrm{fit}_\mathrm{2level}$ by the dashed red line in Figures 2 and 3. \\


\begin{thebibliography}{10}
\bibitem{Meschede2010}
\bibinfo{note}{During the final stage of manuscript preparation we learned
  about related experiments in the group of D. Meschede}.
\end{thebibliography}

\noindent\textbf{Acknowledgements} We thank D. L. Moehring, H. P. Specht, C. N{\"o}lleke, A. Neuzner and C. Guhl for their contributions during the early stages of the experiment. This work was supported by the Deutsche Forschungsgemeinschaft [Research Unit 635] and the European Union [IST programs SCALA and AQUTE]. E. F. acknowledges support from the Alexander von Humboldt Foundation. C. J. V. B. acknowledges support from Coordena\c{c}\~{a}o de Aperfei\c{c}oamento de Pessoal de N\'{i}vel Superior.\\

\end{document}